# Software Requirements Specification of the IUfA's UUIS -- a Team 2 COMP5541-W10 Project Approach

**Omer Shahid Ahmad**

**Faisal Alrashdi**

**Jason (Jun-Duo) Chen**

**Najah Ilham**

**Jianhai Lu**

**Yiwei Sun**

**Tong Wang**

**Yongxin Zhu**

## Table of Contents [1]

## 1. Introduction

The current system was commissioned by the Imaginary University of Arctica (IUfA) in order to efficiently manage the inventory of the university and all requests relevant to such a system. In particular, we were requested to ensure that a central location holding the records of every asset purchased by the IUfA would be accessible through the internet via a compatible browser. Authorized members of the IUfA would be able to view the holdings of the appropriate affiliation (department, faculty or the entire university, according to the user's permission level) and submit requests for a purchase or a transfer (for instance, a transfer of location, or to borrow an item, or even a transfer of ownership). Such requests would be visible to authorized personnel, and the status of the request would be visible to the member who submitted it.

### 1.1 Purpose of this Document

This SRS describes the functional and performance requirements expected from the Unified University Inventory System (UUIS) of the IUfA that will be developed by Team 2. The current document will serve to communicate the project as we understand it to the client, and will also serve as a basis for future reference in case any feature is to be modified or new features are requested in the future.

### 1.2 Scope of the Development Project

The UUIS will hold a record of all items owned by IUfA. Through the web-based application we will develop, IUfA members will be able to view the University holdings relevant to their professional titles (hereafter simply "title") and affiliation, including but not limited to software titles, laboratory materials and computer-related hardware.

This system will decrease the cost of maintaining separate inventory systems as the different locations across the campus will be consolidated. We are not responsible for offering access to the inventory in any language other than English.

### 1.3 Definitions, Acronyms, and Abbreviations

Behaviour Model: An operational principle for all requirements analysis methods. A state transition diagram represents the behaviour of a system by depicting its states and the events that cause the system to change state [8].

Chrome: A web browser developed by Google Inc., which can run on Microsoft Windows, Apple Mac OS X and Linux.

CAPTCHA: A system intended to websites against bots by generating and validating tests that humans are expected pass but current computer programs are not. The term CAPTCHA, which stands for "Completely Automated Public Turing Test to Tell Computers and Humans Apart", was coined in 2000 by Luis von Ahn, Manuel Blum, Nicholas Hopper and John Langford of Carnegie Mellon University [17].

Data Dictionary: An organized listing of all data elements that are pertinent to the system with precise and rigorous definitions, so that both users and system analysts will have a common understanding of input, output, components of stores and intermediate calculations [8].

Domain Model: Model documenting key concepts, and the domain-vocabulary of the system being modeled. The model identifies the relationships among all major entities within the system, and usually identifies their important methods and attributes. The domain model provides a structural view of the system that can be complemented by other dynamic views in Use Case models [2].

Firefox: A free and open source web browser descended from the Mozilla Application Suite and managed by Mozilla Corporation. It runs on Microsoft Windows, Apple Mac OS X and Linux [11].

Functional Requirements: Defines the inputs, behaviour, and outputs of a software system or its components. This may be achieved via calculations, technical details, data manipulation and processing and/ other specific functionality that define what a system is expected to accomplish [12].

HTTPS: "Hypertext Transfer Protocol Secure", a combination of the Hypertext Transfer Protocol with the SSL/TLS protocol to provide encryption and secure identification of the server.

Internet Explorer: Windows Internet Explorer (formerly Microsoft Internet Explorer; abbreviated to MSIE or, more commonly, IE), is a series of graphical web browsers developed by Microsoft Corporation. It runs on Microsoft Windows, Windows CE, Windows Mobile, Apple Mac OS and UNIX [13].

iPhone: A line of Internet and multimedia-enabled smartphone designed and marketed by Apple Inc.[14]

IUfA: Imaginary University of Arctica.

Non-Functional Requirements: The requirements specifying the criteria that can be used to judge the operation of a system, rather than specific behaviours. This should be contrasted with functional requirements that define specific behaviour or functions. [14]

Role: Combination of the set of permissions and affiliation granted to a user for a title adopted by the user. Please refer to the definition of "title" for disambiguation.

Safari: A web browser developed by Apple. It runs on Apple Mac OS, iPhone OS and Microsoft Windows.

SRS: Software Requirements Specification.

Smartphone: A mobile phone offering advanced capabilities, often with PC-like functionality (PC-mobile handset convergence). There is no industry standard definition of a smartphone [19].

Software Title: A given version of a software solution or a set of software solutions released in a single bundle.

SSL/TLS: Transport Layer Security (TLS) and its predecessor, Secure Sockets Layer (SSL), are cryptographic protocols that provide security for communications over networks such as the Internet. TLS and SSL encrypt the segments of network connections at the Transport Layer end-to-end.[18]

Title: The professional title of an IUFA member, e.g. the position held by the member with regards to the university.

Use Cases: The correct behaviour for the situation described. Use cases are not functions or features and they cannot be decomposed. Use cases have a name and a brief description. They also have detailed descriptions that are essentially stories about how the users use the system to do something they consider important, and what the system does to satisfy these needs.[16]

UUIS: Unified University Inventory System.

### 1.4 References

[1] SRS template. [online access, Feb 20, 2010]. 2010. http://www.sju.edu/~scooper/spring01se/SoftwareRequirementsSpecification.htm
[2] Wikipedia. Domain model | Wikipedia, the free encyclopaedia. [Online; accessed 2010-02-20]. http:http://en.wikipedia.org/wiki/Domain_model
[3] Domain Modelling | Paul Oldfield, . [Online; accessed 2010-02-20]. Mentors of Cally Ltd. http://www.aptprocess.com/whitepapers/DomainModelling.pdf
[4] Object Thinking Domain Model Example | dotNetTemplar . [Online; accessed 2010-02-21]. http://dotnettemplar.net/Object+Thinking+Domain+Model+Example.aspx
[5] A Quick Guide to The Unified Modeling Language (UML) | California State University, San Bernardino. [Online; accessed 2010-02-21]. http://www.csci.csusb.edu/dick/samples/uml0.html#Domain.
[6] UML 2 Class Diagram Guidelines | Scott W. Ambler, Ambysoft Inc. [Online; accessed 2010-02-21]. http://www.agilemodeling.com/style/classDiagram.htm
[7] UML Tutorial: Part 1 -- Class Diagrams.|Robert C. Martin [Online; accessed 2010-02-21]. http://www.objectmentor.com/resources/articles/umlClassDiagrams.pdf
[8] Software Engineering - A Practitioner's Approach 5th Edition | Roger S. Pressman, McGraw Hill, 2001
[9] Wikipedia. BlackBerry | Wikipedia, the free encyclopedia. [Online; accessed 2010-02-22]. http://en.wikipedia.org/wiki/BlackBerry
[10] Wikipedia. Google Chrome | Wikipedia, the free encyclopedia. [Online; accessed 2010-02-22]. http://en.wikipedia.org/wiki/Google_Chrome
[11] Wikipedia. Mozilla Firefox | Wikipedia, the free encyclopedia. [Online; accessed 2010-02-22]. http://en.wikipedia.org/wiki/Mozilla_Firefox
[12] Wikipedia. Functional requirement | Wikipedia, the free encyclopedia. [Online; accessed 2010-02-22]. http://en.wikipedia.org/wiki/Functional_requirement
[13] Wikipedia. Internet Explorer | Wikipedia, the free encyclopedia. [Online; accessed 2010-02-22]. http://en.wikipedia.org/wiki/Internet_Explorer
[14] Wikipedia. Non-functional requirement | Wikipedia, the free encyclopedia. [Online; accessed 2010-02-22]. http://en.wikipedia.org/wiki/Internet_Explorer
[15] Wikipedia. Safari (web browser) | Wikipedia, the free encyclopedia. [Online; accessed 2010-02-22]. http://en.wikipedia.org/wiki/Safari(web_browser)

[16] Use Case Modelling by Kurt Bittner, Ian Spence, Ivar Jacobson | Pearson Education, Inc.
[17] Carnegie Mellon University. CAPTCHA | Carnegie Mellon University [Online; accessed 2010-02-23]. http://www.captcha.net/
[18] Wikipedia. Transport Layer Security | Wikipedia, the free encyclopedia. [Online; accessed 2010-02-22]. http://en.wikipedia.org/wiki/Transport_Layer_Security
[19] ] Wikipedia. Smartphone | Wikipedia, the free encyclopedia. [Online; accessed 2010-02-23]. http://en.wikipedia.org/wiki/Smartphone
[20] Pfleeger, S. L. & Atlee, J. M. (2010). Software Engineering: Theory and Practice (4th Ed.). Prentice Hall, Pearson Higher Education: Upper Saddle River.
[21] Serguei Mokhov. COMP5541_Tools and Techniques for Software Engineering, Assignment 2. [Online; accessed 2010-02-23]. http://users.encs.concordia.ca/~c55414/ta2.pdf
[22] Hossein Kassaei, Jeff Miles, Lei Chen,Li Qun Sun,Sameh Khairalla,Weidong Bao,Xiaodan Wu,Zaki Saad Zaki, FAMA SRS.doc [online; accessed Mar-06-2010], 2007. http://users.encs.concordia.ca/~c55414/samples/comp5541-f07/team1/FAMA%20SRS.doc
[23]EstimationGuidelines. [online access 10-03-2010]. http://www.projects.ed.ac.uk/methodologies/Full_Software_Project_Template/EstimationGuidelines.shtml

## 1.5 List of Tables and Figures

### 1.5.1 Tables:

**Figures:**

## 1.6 Overview of Document

This document contains a general description of the product, functional requirements, non-functional requirements, and use cases analyses as well as a behaviour model, data dictionary and domain model.

## 2. General Description

This unified inventory system for all the Faculties of IUfA can do the management of inventory of their various assets (equipment, furniture, space, software, seat assignment, etc). Users may log in to the system. Also, users can submit a request to inventory or report a problem with an inventory item with or without a barcode, serial number, and/or a description (level 0). Changes are made based on the request and submitted for approval to become permanent.

The UUIS has three levels of approval:

1. Technical staff (IT and techies) (level 1).
2. DA (level 2).
3. Chair or director FA, e.g. Associate Dean, Dean, Controller (level 3).

The users may log out from the system at any time during the session.

### 2.1 User Characteristics

The users are expected to have basic computer knowledge, such as prior experience in logging into a system and viewing inventory data.

### 2.2 Product Perspective

This product currently aims to consolidate the inventory listings from the entire university into a single system.

### 2.3 Overview of Functional Requirements

Users must first be authenticated to logon to and log out from the site. Once a user gets into the site, they may do some request operations (submit, view, and cancel). In addition, they can search for certain information of assets and print out the search results.

High levels users can also do some management jobs, such as University Management (bulk import users, data base backup), Error Management (list, print & view details), Request Management (view, approve & reject).

Properly authorized users may audit the database. They can do the Logging and auditing (audit information should be browsable as if it were from regular data).

**2.4 Overview of Data Requirements**

The user must first authenticate their identity by providing a valid combination of username, or userID, and password. The user must then choose to (1) browse the inventory (provided the user has the appropriate permission level), (2) review account details (such as past transactions, or status of ongoing transactions, or personal information), or (3) log out. The items viewable must correspond to the permission level of the user.

Therefore, the system must contain a listing of the items as well as their relevant characteristics (the department owning the item, the physical location of the item, properties of the item, barcodes for all such items, etc.), the members of the IUfA (including related information such as user ID, password, department, role within the department, etc.) and the details of each transaction.

In addition, the system must keep a record of all critical changes to the system. This is to allow properly authorized users to audit the database, and verify the consistency and correctness of the database as well as trace the history of items should any errors arise.

### 2.5 General Constraints, Assumptions, Dependencies, Guidelines

The product must be Web-based. We assume that users will have some familiarity with Web-based inventory systems.

## 3 Functional Requirements

- Authorized users must be able to login with their identification number (or username) and their password.
- After a successful login, the user may either search for items by inventory or browse through a list of the items. Users may only view items allowed by their permission level.
- After the user login, he/she may do some operations of request, like submit a request, view the request, or cancel the request.
- The user with certain permission level can also do asset management jobs (view, add, update and bulk add assets).
- The user with certain permission level can do university management jobs, such as to create a department, to create a faculty, to add a location, and bulk import users from a CSV file.
- The user with certain permission level can review assets (view audit options, audit logs, produce report as well as output review).
- The user with certain permission level can also do error management (list error messages, print error messages and View more details/annotate error messages).
- Properly authorized users may audit the database. During an audit, the history of all transactions of any database object (user, location, department, item, etc.) may be viewed, and double-checked with the data obtained from other objects for internal consistency.

## 4. Non-Functional Requirements

The system should meet the following non-functional requirements.

### 4.1 Usability

The web interface of the system will be designed to be concise and user-friendly, with a graphical interface to help users identify the proper choice on the screen. An online help will be provided for the users. Users are expected to be able to use the system productively with minimal or no training. We will include legacy inventory records in order to help users understand the new system of organization.

### 4.2 Reliability and Robustness

The system should be available at all times (24 hours a day, 7 days a week) except for monthly maintenance of no more than 10 minutes. The system backup onto a second server will be performed during the maintenance time as well as once daily (at midnight) and system recovery will only be executed as necessary. In addition, the secondary server will be prepared to help recover in the event of hardware failure.

### 4.3 Verifiability

All critical transactions will be recorded in a separate table in order to allow for auditing the database system, such that all suspicious records may be easily tracked and the appropriate record double-checked as necessary. The audits must be performed by an authorized user, with special privileges granted to allow visibility of the auditing function and associated records in the database.

### 4.4 Performance

All pages should be loaded within three seconds or less, assuming a broadband connection on the client side. Therefore, response time for transactions will be three seconds or less. The system should support as many as 50 online users simultaneously with negligible response delay.

### 4.5 Portability and Interoperability

As a web-based application, the system will support the latest version of the majority of browsers such as Internet Explorer, Firefox, Chrome and Safari, as well as common mobile devices (Blackberry, iPhone, etc.). In addition, the system should be easily migrated to other platform in case of hardware failure in both servers.

### 4.6 Security

Only authorized users will be permitted to access the system. The system will provide additional security means to protect itself from automated attacks by using methods such as "CAPTCHA" when processing login requests in special cases (when the user has elevated privileges, or when a user repeatedly attempts to login without success). The system will provide the users with a secure way to change their passwords, whether when initializing a new account or by user request. If requested, we will upgrade the system to use the HTTPS protocol in subsequent iterations in order to prevent unauthorized third-party viewing of the contents, though the university would be responsible for providing the proper certificates in such a scenario.

### 4.7 Maintainability

The standardized design and implementation documents will be provided in order to maintain the system. All changes will be documented. A standard architecture will be applied and therefore the system should be easy expandable, allowing for quick evolution of the software to adapt to possible situations in the future. In addition, these documents should allow third parties to be consulted should the inventory system exhibit abnormal behaviour.

## 5. Use Cases and Behaviour Model [22]

In this section, we will describe all the use case scenarios and the interaction of user in tables. The purpose for this is for developers as a guideline when developing user interface of system [1]. The following represents the requirements, the fundamental functions that an inventory system should have.

### 5.1.1 Use Case to Login

| ***Login*** | |
|---|---|
| **Description** | Authenticates user and grants access to system. |
| **Actors** | All users |
| **Trigger** | User clicks the "Login" button from the home page |
| **Pre-conditions** | 1. User is not yet logged in.<br>2. User is subscribed to the UUIS. |
| **Post-conditions** | User accesses the interface corresponding to his/her role |
| **Main Case** | 1. User enters his/her ID or username and password and launches the login process.<br>2. System verifies whether the username/password pair is valid.<br>3. System displays the "Welcome" page, tailored to user's permissions. |
| **Exception Path** | 1. User enters an invalid username/password combination.<br>2. System denies access to user.<br>3. System displays error message and prompts user to retry.<br>4. User may repeat login procedure.<br>5. User is locked out after the third attempt. |

Table 5.1.1 Common Use Case to Login.

### 5.1.2 Use Case to Logout

| ***Logout*** | |
|---|---|
| **Description** | Exits user from system. |
| **Actors** | All users |
| **Triggers** | User clicks the logout button, or connection times out. |
| **Pre-conditions** | User is already logged in. |
| **Post-conditions** | User is directed to the login page |
| **Main Case** | 1. User requests to log out from system or connection times out.<br>2. System takes back access from user.<br>3. The login page is displayed. |
| **Exception Path** | 1. User requests to log out from system or connection times out.<br>2. System displays an error message.<br>3. The logout is not completed.<br>4- User must retry to logout. |

Table 5.1.2 Common Use Case to Logout. **Note:** this function is available in all pages. In addition, system will automatically log out user after 30 minutes of inactivity.

### 5.1.3 Use Case to Change Password

| ***Change Password*** | |
|---|---|
| **Description** | Updates user's password. |
| **Actors** | All users |
| **Trigger** | User clicks the "change password" button |
| **Pre-conditions** | User is logged in |
| **Post-conditions** | 1. Viewing "password successfully updated "page<br>2. Password updated |
| **Main Case** | 1. System displays "change password" page.<br>2. System prompts user to enter the old password once and the new password twice.<br>3. User enters old and new passwords.<br>4. User selects the submit button.<br>5. System verifies old and new passwords.<br>6. User's new password is saved. |
| **Exception Paths** | 1. User's old password is incorrect, or new passwords do not match.<br>2. System detects the error before attempting the update.<br>4. System displays an error message.<br>5. User may repeat the procedure. |

Table 5.1.3 Use Case to Change Password.

### 5.1.4 Use Case to View/Edit Personal Information

| ***View/Edit Personal Information*** | |
|---|---|
| **Description** | Displays and allows editing of user information. |
| **Actors** | All users |
| **Trigger** | User Clicks "My Profile" from the menu. |
| **Pre-conditions** | User is logged in. |
| **Post-conditions** | 1. Viewing updated personal information.<br>2. Database is updated. |
| **Main Case** | 1. User chooses to view/update his/her profile.<br>2. System retrieves the appropriate information from the database.<br>3. System displays the appropriate information to user, along with the option of updating the allowed fields.<br>4. User views the information, and updates the information where relevant.<br>5. System updates the information if requested by user.<br>6. User exits the page. |
| | 7. System commits the changes to the database. |
| **Exception Path** | 1. Database error<br>2. System displays error message and with an apology. |

Table 5.1.4 Use Case to View/Edit Personal Information.

### 5.1.5 Use Case to Submit a Request

| ***Submit a Request*** | |
|---|---|
| **Description** | Submits a new request. |
| **Actors** | Users with all permission levels |
| **Trigger** | User selects the "Submit new request" option. |
| **Pre-conditions** | User is already logged in. |
| **Post-conditions** | 1. User is viewing the request page.<br>2. New request stored in the database |
| **Main Case** | 1. System displays the "Request" form with all the options available to user.<br>2. User enters the relevant asset identification numbers (serial number, location ID, etc., as allowed according to user's permission level) along with a description of the request (type of request, reason for request, etc.).<br>3. User selects the submit button.<br>4. System displays a message including the relevant information entered by user to seek confirmation of the request.<br>5. User confirms the request<br>6. System commits the request to the database<br>7. System displays a "request submitted successfully" message. |
| **Exception Path** | 1. User cancels the request<br>2. System does not proceed with the request and clears the fields on the request page, but does not move to another page.<br>3. A database error occurs before committing the request to the database.<br>4. System logs the error, and saves the request information along with the details of the error.<br>5. System displays an error message requesting user to contact a system administrator. |

Table 5.1.5 Use Case to Submit a Request.

### 5.1.6 Use Case to View Request Status

| ***View Request Status*** | |
|---|---|
| **Description** | Displays status of user's requests. |
| **Actors** | All users |
| **Trigger** | User Clicks "Request Status" button. |
| **Pre-conditions** | User is logged in, and user has previously submitted requests. |
| **Post-conditions** | 1. Viewing updated personal information<br>2. Database is updated |
| **Main Case** | 1. User selects the option of viewing request status.<br>2. System displays the current status for the request(s) made by user. |
| **Exception Path** | 1. No requests have been recorded in system.<br>2. System displays a message indicating that user has made no requests. |

Table 5.1.6 Use Case to View Request Status. **Note:** Users may submit as many requests as they wish. This function displays a list of all the requests that one user has made.

### 5.1.7 Use Case to Cancel a Request

| ***Cancel a Request*** | |
|---|---|
| **Description** | Cancels user's pending request(s). |
| **Actors** | All users |
| **Trigger** | User Clicks "Cancel Request " button. |
| **Pre-conditions** | 1. User is logged in.<br>2. User is able to view his/her pending requests. |
| **Post-conditions** | 1. User confirms the cancellation of the request.<br>2. Request(s) are cancelled.<br>3. User is returned to view the status of his/her requests. |
| **Main Case** | 1. System displays the request(s) made by user<br>2. User selects the request(s) he/she wishes to cancel<br>. System displays a message including the request(s) to be cancelled and prompts user to confirm<br>4. User confirms the cancellation<br>. System updates the status of the request(s) and displays a "request(s) cancelled successfully" message.<br>6. System returns user to view the status of his/her requests after a timed delay. |
| **Exception Path** | 1. Database error<br>2. System displays an error message<br>3. User chooses to cancel the "request cancellation"<br>4. System displays the "request cancellation" page |

Table 5.1.7 Use Case to Cancel a Request.

### 5.1.8 Use Case to Search for Data

#### 5.1.8.1 Use Case to Search for Data

| ***Search*** | |
|---|---|
| **Description** | Searches for data. |
| **Actors** | All users |
| **Trigger** | User Clicks "Search" button. |
| **Pre-conditions** | 1. User is logged in. |
| **Post-conditions** | 1. User views the data conforming to the search parameters.<br>2. User is returned to the previous view without having searched for data. |
| **Main Case** | 1. System displays the simple search interface.<br><br>Case I.<br>2. User enters a simple query and initiates the search.<br>3. System parses the request and performs the query.<br>4. System displays the data retrieved by the query.<br><br>Case II.<br>2. User requests the option of an advanced search<br>3. System presents the interface for an advanced search.<br>4. User either cancels the query (Case III), or user enters the search parameters.<br>5. System parses the request and performs the query.<br>6. System displays the data retrieved by the query.<br><br>Case III.<br>2. User cancels the search request.<br>3. System removes the search interface, returning to the page viewed by user prior requesting for a search. |
| **Exception Path** | 1. Database error<br>2. System displays an error message<br>3. User chooses to cancel the "request cancellation"<br>4. System displays the "request cancellation" page |

Table 5.1.8.1 Search for Data.

5.1.8.2 Print/Save Search Results

| ***Print/Save Search Results*** | |
|---|---|
| **Description** | Outputs search results. |
| **Actors** | All users |
| **Trigger** | User Clicks "Print/save" button. |
| **Pre-conditions** | 1. User is logged in.<br>2. User has the search result page open. |
| **Post-conditions** | 1. The search is saved and/or printed.<br>2. User is returned to the search page. |
| **Main Case** | 1. System displays the search results.<br>2. User selects the data he/she finds of interest.<br>3. User clicks the print or save button.<br>4. System displays a message requesting the confirmation of user along with the number of entries selected.<br>5. User confirms the action requested.<br>6. System outputs the information as requested. |
| **Exception Path** | 1. User does not select data.<br>2. System selects all search results by default.<br>3. System proceeds to request confirmation (Main case, #4). |

Table 5.1.8.2 Exporting Search Results.

## 5.2 Use Cases for University Management

### 5.2.1 Use Case to Create a Department

| ***Create Department*** | |
|---|---|
| **Description** | Creates a new Department. |
| **Actors** | Users with permission level 2 |
| **Trigger** | Admin user clicks the "Add Department" option from the menu |
| **Pre-conditions** | 1. Authenticated session. |
| **Post-conditions** | 1. The Department is created.<br>2. User is returned to the "University Management" page. |
| **Main Case** | 1. User clicks on the "Add Department" option from the menu.<br>2. User edits fields as desired in the page<br>3. User selects the "submit" button<br>4. System requires confirmation from the faculty head in the form of the faculty Dean's username and password.<br>6. User confirms the modification along with the faculty head.<br>7. System updates the database and returns user to the "Manage Roles" page. |
| **Exception Path** | 1.1. User cancels the creation of a Department.<br>1.2. System does not save the information entered and reloads the "Manage roles" page.<br>2.1. Incomplete information in the page.<br>2.2. System displays an error message.<br>2.3. User is given another opportunity to re-edit the form.<br>3.1. Database error.<br>3.2. System logs the error along with the information on the modifications.<br>3.3. System notifies user of unsuccessful update and requests user to contact a system administrator. |

Table 5.2.1 Use Case to Create a Department.

## 5.2.2 Use Case to Create a Faculty

| ***Create Faculty*** | |
|---|---|
| **Description** | Creates a new Faculty. |
| **Actors** | Users with permission level 3 |
| **Trigger** | Admin user clicks the "Add Faculty" option. |
| **Pre-conditions** | Authenticated session. |
| **Post-conditions** | 1. The Faculty is created.<br>2. User is returned to the "University Management" page. |
| **Main Case** | 1. User clicks on the "Add Faculty" option from the menu.<br>2. User edits fields as desired in the page<br>3. User selects the "submit" button<br>4. Unless user is the principal, system requires confirmation from the principal in the form of the principal's username and password. If user is the principal, then system only requires confirmation from user.<br>6. User confirms the modification along with the principal.<br>7. System updates the database and returns user to the "University Management" page. |
| **Exception Path** | 1.1. User cancels the creation of a Faculty.<br>1.2. System does not save the information entered and reloads the "University Management" page.<br>2.1. Incomplete information in the page.<br>2.2. System displays an error message.<br>2.3. User is given another opportunity to re-edit the form.<br>3.1. Database error.<br>3.2. System logs the error along with the information on the modifications.<br>3.3. System notifies user of unsuccessful update and requests user to contact a system administrator. |

Table 5.2.2 Use Case to Create a Faculty.

## 5.2.3 Use Case to Add a Location

| ***Add a Location*** | |
|---|---|
| **Description** | Creates new location in database. |
| **Actors** | Users with permission level 2 or 3 |
| **Trigger** | Admin user clicks the "Add location" option from the "University Management" menu or from the "University Management" page. |
| **Pre-conditions** | Authenticated session. |
| **Post-conditions** | 1. The location is added<br>2. User is returned to the "University Management" page. |
| **Main Case** | 1. User clicks on the "Add Location" option from the menu, or from the "University Management" page.<br>2. System presents user with the appropriate form, with fields filled-in according to user's role profile.<br>3. User completes the fields and selects the "submit" button<br>4. System requests confirmation.<br>6. User confirms the modification.<br>7. System updates the database and returns user to the "University Management" page. |
| **Exception Path** | 1.1. User cancels the addition of a location.<br>1.2. System does not save the information entered and reloads the "University Management" page.<br><br>2.1. Incomplete information in the page.<br>2.2. System displays an error message.<br>2.3. User is given another opportunity to re-edit the form.<br><br>3.1. Database error.<br>3.2. System logs the error along with the information on the modifications.<br>3.3. System notifies user of unsuccessful update and requests user to contact a system administrator. |

Table 5.2.3 Add a Location.

## 5.2.4 Use Case to Back up Data in Database

| ***Back Up Data*** | |
|---|---|
| **Description** | Creates back-up of data. |
| **Actors** | Users with permission level 1, 2 or 3 |
| **Trigger** | User clicks the "Back-up" option from the menu |
| **Pre-conditions** | 1. User is logged in. |
| **Post-conditions** | 1. Information is backed up.<br>2. User receives confirmation. |
| **Main Case** | 1. User selects "Back-up" option.<br>2. System displays options for back-up: user-related information, university-related information, inventory-related information, request-related information or entire database.<br>4. User selects the appropriate option and launches the back-up.<br>5. System asks for confirmation, along with a warning that the operation may take a few minutes and cause a slowdown of system for all users.<br>6. User confirms.<br>7. System exports the requested data to CSV files.<br>8. System compares CSV files to database records. System notifies user of successful back-up if the two are identical, or displays the number of records which are different if there are any. |
| **Exception Path** | 1.1. User cancels the operation<br>1.2. System does not proceed with the back-up and returns user to the "Welcome" page. |

Table 5.2.4 Use Case to Back up Bata in Database. **Note**: Back-ups will be automatically created once per week. However, users with sufficient privileges will be able to manually back-up the data at any time (for instance, after bulk adding users).

### 5.2.5 Use Case to Bulk Import Users from a CSV File

| ***Import Users*** | |
|---|---|
| **Description** | Bulk adds users from CSV file. |
| **Actors** | Users with permission level 1, 2 or 3 |
| **Trigger** | Admin user selects the "Import users" option from the menu |
| **Pre-conditions** | 1. User is logged in. |
| **Post-conditions** | 1. Information about new users is imported.<br>2. User is returned to the "Users" page. |
| **Main Case** | 1. User selects "import users" option.<br>2. User enters required information: file name, path, etc.<br>3. User selects "Submit" button.<br>4. System asks for confirmation.<br>5. User confirms.<br>6. System adds the appropriate record and returns user to the "Users" page. |
| **Exception Path** | 1.1. User cancels the operation.<br>1.2. System does not proceed with adding and updates "User information" page.<br><br>2.1. Incomplete information in the page.<br>2.2. System displays an error message.<br>2.3. User is informed that the operation was unsuccessful.<br>2.4. User is returned to the "Users" page. |

Table 5.2.5 Use Case to Import User from a CSV File.

### 5.2.6 Use Case to Update Target User's Role Profile

| ***Update User Profile*** | |
|---|---|
| **Description** | Updates a user's profile |
| **Actors** | Users with permission levels 1, 2 or 3 |
| **Trigger** | User clicks the "modify" option for a given user. |
| **Pre-conditions** | 1. User is logged in.<br>2. User is viewing the intended user's information.<br>3. User has selected the specific role to be modified. |
| **Post-conditions** | 1. Role is updated.<br>2. User is returned to the "Users" page. |
| **Main Case** | 1. System displays the "user information" page.<br>2. User edits the permission (or other such fields, if applicable) as required.<br>3. User selects the "submit" button.<br>4. System asks for confirmation.<br>5. User confirms the modification.<br>6. System commits the modifications and returns user to the "Users" page. |
| **Exception Path** | 1.1. User cancels the operation.<br>1.2. System returns user to "Users" page without exporting the data.<br><br>2.1. Database error occurs.<br>2.2. System logs the error along with the record(s).<br>2.3. User is informed that the operation was unsuccessful.<br><br>3.1. Required fields are left blank, or filled with invalid data.<br>3.2. System notifies user of the problem and returns user to user profile update page. |

Table 5.2.6 Use Case to Update Target User's Profile. **Note:** an admin user may only edit the role profile of a target user whose permission level is lower than that of the admin user.

## 5.3 Use Cases for Asset Management

### 5.3.1 Use Case to View Assets

| ***View Assets*** | |
|---|---|
| **Description** | Displays assets to user, filtered by the user's permissions. |
| **Actors** | Users with permission level 1, 2 or 3 |
| **Trigger** | User selects the "View Assets" option from the menu. |
| **Pre-conditions** | User is logged in. |
| **Post-conditions** | User views a list of assets available to user. |
| **Main Case** | 1. System retrieves assets information<br>2. System displays the "assets information" page with the appropriate information |
| **Exception Path** | 1. A database error occurs.<br>2. System logs the error.<br>3. System displays an error message and an apology. |

Table 5.3.1 Use Case to View Assets. Note: The resulting view is different for different users having access to this function as a result of their role and permission level in system.

### 5.3.2 Use Case to Add Assets

| *Add Assets* | |
|---|---|
| **Description** | Adds a single asset. |
| **Actors** | Users with permission level 1, 2 or 3 |
| **Trigger** | User selects the "Add assets" option from the menu. |
| **Pre-conditions** | 1. User is logged in.<br>2. User has opened the "Assets" menu. |
| **Post-conditions** | 1. Asset is created.<br>2. User is returned to the "View Assets" page. |
| **Main Case** | 1. User opens the "Assets" menu and chooses to add assets.<br>2. User enters required information (asset name, barcode, location(s), etc.)<br>3. User selects "Add assets" button.<br>4. System requests confirmation.<br>5. User confirms adding the asset.<br>6. System updates "Manage assets" page. |
| **Exception Path** | 1.1. User cancels the operation.<br>1.2. System does not proceed with adding the asset and user is returned to the "View Assets" page.<br><br>2.1. Required field(s) incomplete or incorrect<br>2.2. System informs user of the problems and returns user to the "Add Assets" page.<br><br>3.1. Database error.<br>3.2. System logs the error along with the completed fields.<br>3.3. System displays an error message informing user of an unsuccessful inventory update.<br>3.4. User is returned to the "View Assets" page. |

Table 5.3.2 Use Case to Add Assets.

### 5.3.3 Use Case to Update Asset Information

| *Update Asset Information* | |
|---|---|
| **Description** | Updates information of existing asset(s). |
| **Actors** | Users with permission level 1, 2 or 3 |
| **Trigger** | User selects asset(s) and clicking the "Modify asset(s)" button |
| **Pre-conditions** | 1. User is logged in.<br>2. User is able to view assets. |
| **Post-conditions** | 1. Database updated.<br>2. "Asset Updated" page is displayed. |
| **Main Case** | 1. User selects the asset(s) of interest and clicks on "modify".<br>2. System displays the "modify asset" page.<br>3. User edits fields as desired in the page.<br>4. User selects the "submit" button.<br>5. System asks for confirmation.<br>6. User confirms the modification.<br>7. System updates the database.<br>8. System returns user to the "View Assets" page. |
| **Exception Path** | 1.1. User cancels the modification.<br>1.2. System does not proceed with modification, and reloads the "Manage assets" page.<br><br>2.1. User does not select any asset to modify.<br>2.2. System informs user of error and returns to the "Assets" page.<br><br>3.1. Required fields are incomplete.<br>3.2. System displays an error message.<br>3.3. User is returned to the form to complete the required fields.<br><br>4.1. Database error before system commits the changes.<br>4.2. System logs the error along with the changes requested.<br>4.3. System displays an error message indicating that the update was not successful. |

Table 5.3.3 Use Case to Update Asset Information.

### 5.3.4 Use Case to Bulk Add Assets

| ***Bulk Add Assets*** | |
|---|---|
| **Description** | Adds a large number of assets. |
| **Actors** | Users with permission level 1, 2 or 3 |
| **Trigger** | User clicks the "Add assets" button. |
| **Pre-conditions** | 1. User is logged in.<br>2. User has opened the "Assets" menu. |
| **Post-conditions** | 1. Asset is created.<br>2. User is returned to the "View Assets" page. |
| **Main Case** | 1. User opens the "Assets" menu and chooses to add assets.<br>2. User uploads a single CSV file containing all of the relevant information for each item. The first line of the CSV file (the "header entry") will contain the fields for all subsequent entries.<br>3. User selects "Add assets" button.<br>4. System parses the file to ensure that the header entry is acceptable. System then verifies that all subsequent entries are entered in a compatible format.<br>5. System requests confirmation.<br>6. User confirms adding the asset.<br>7. System updates system.<br>8. System displays the assets, pre-selected for user to modify properties if necessary. |
| **Exception Path** | 1.1. User cancels the operation.<br>1.2. System does not proceed with adding the asset and user is returned to the "View Assets" page.<br><br>2.1. CSV file contains internal errors (e.g. the header entry is incorrect, or the subsequent entries do not match the header entry).<br>2.2. System informs user of the problems and returns user to the "Add Assets" page.<br><br>3.1. Database error.<br>3.2. System logs the error along with the completed fields.<br>3.3. System displays an error message informing user of an unsuccessful inventory update.<br>3.4. User is returned to the "View Assets" page. |

Table 5.3.4 Use Case to Bulk Add Assets.

### 5.3.5 Group Assets

| ***Group Assets*** | |
|---|---|
| **Description** | Groups assets into logical clusters. |
| **Actors** | Users with permission level 1, 2 or 3 |
| **Trigger** | User clicks the "Group" button. |
| **Pre-conditions** | 1. User is logged in.<br>2. User has opened the "Assets" menu. |
| **Post-conditions** | 1. Asset is created.<br>2. User is returned to the "Assets" page. |
| **Main Case** | 1. User selects assets to be grouped.<br>2. User clicks on the "Group assets" button.<br>3. System verifies that none of the assets were previously grouped.<br>4. System updates the database to group the items together.<br>5. System returns user to the "Assets" page, along with a confirmation that the assets have been grouped. |
| **Exception Path** | 1.1. The assets selected have been previously grouped together, and are the only assets in the group.<br>1.2. System does not proceed with grouping the assets a second time and notifies user. User is returned to the "Assets" page.<br><br>2.1. The assets selected have been previously grouped together, and other asset(s) (viewable by user) is (are) also in the group but not selected.<br>2.2. System requests confirmation for removing the item(s) from the group.<br>2.3. User confirms removing the asset(s) from the group.<br>2.4. System updates the database to remove the asset(s) from the group. (Note: It is also possible to remove an asset from a group by updating the asset properties.)<br><br>3.1. The assets selected have been previously grouped together, and other asset(s) not viewable by user is (are) also in the group but not selected.<br>3.2. System notifies user to request a higher-level user to perform the removal and returns user to the "Assets" page.<br><br>4.1. The assets belong to two or more different groups, and all the items in the different groups are selected.<br>4.2. System requests confirmation for merging the two groups.<br>4.3. User confirms request.<br>4.4. System updates the database and returns user to the "Assets" page.<br><br>5.1. The assets belong to two or more different groups, and some items in at least one group have not been selected.<br>5.2. System requests confirmation for removing the selected items from their current groups and creating a new group.<br>5.3. User confirms request.<br>5.4. System updates the database and returns user to the "Assets" page. |

Table 5.3.5 Group Assets.

## 5.4 Use Cases for Reviewing

### 5.4.1 View Auditing Options

| ***View Audit Options*** | |
|---|---|
| **Description** | Updates a user role profile. |
| **Actors** | Admin users |
| **Trigger** | User clicks the "Audit" option from the menu or from the welcome page. |
| **Pre-conditions** | 1. Authenticated session. |
| **Post-conditions** | 1. Auditing options are displayed |
| **Main Case** | 1. User chooses on the "Audit" option.<br>2. System processes and displays the options available according to user's role profile (asset, time interval, user, department, faculty and/or university-wide). |
| **Exception Path** | 1. Database error.<br>2. System logs the error and requests user to retry. |

Table 5.4.1 View Auditing Options.

### 5.4.2 Audit Logs

| ***Audit by Assets*** | |
|---|---|
| **Description** | Displays a list of transactions. |
| **Actors** | Admin user |
| **Trigger** | User chooses an audit option from the page or from the menu. |
| **Pre-conditions** | 1. Authenticated session. |
| **Post-conditions** | 1. The relevant transactions are listed. |
| **Main Case** | 1. Admin user selects a category (asset, user, department or faculty) from those available in the audit page or in the menu. The options shown are filtered according to user's permission level.<br>2. System compiles a list of all relevant transactions and summarizes the information by asset.<br>3. Admin user requests to view detail.<br>4. System retrieves all records relevant to the asset(s) of interest and displays the information to user. This process may be repeated by expanding each record successively<br>5. User acknowledges the information.<br>6. System returns user to the list of transactions summarized by asset. User may repeat steps 4, 5 and 6 as many times as desired. |
| **Exception Path** | 1. Database error.<br>2. System logs the error along with the information being processed.<br>3. System notifies user of unsuccessful update and requests user to retry. |

Table 5.4.2 Audit Logs.

### 5.4.3 Produce Reports

| ***Produce Reports*** | |
|---|---|
| **Description** | Produces view of the data of interest. |
| **Actors** | Admin users |
| **Trigger** | User selects the "Report" option from either the menu or the "Review" page. |
| **Pre-conditions** | 1. Authenticated session.<br>2. Target information is selected. |
| **Post-conditions** | 1. Target information is displayed |
| **Main Case** | 1. Admin user clicks on the "Report" option.<br>2. System displays a list of possible types of information available to admin user (locations, assets, users, etc.).<br>3. Admin user selects the option of interest.<br>4. System lists all items within admin user's scope corresponding to the admin user's selection.<br>5. Admin user selects the item(s) of interest.<br>6. System lists the options available to admin user for reporting (properties of the item(s) of interest, comparison to other item(s) or comparison within the list of items).<br>7. Admin user repeats steps 4 to 6 (inclusive) until admin user clicks the "Report" button.<br>8. System compiles the information as requested and filters the data according to admin user's permissions. System displays the data as requested. |
| **Exception Path** | 1.1. User did not select any information.<br>1.2. System selects all the information on display and proceeds with the request.<br><br>21. Admin user does not have sufficient privilege to view any of the information requested.<br>2.2. System notifies admin user of the error and returns admin user to the "Report" page after a delay of ten seconds. |

Table 5.4.3 Produce Report.

### 5.4.4 Print/Save Review

| ***Print/Save Audit Logs*** | |
|---|---|
| **Description** | Records the information of interest. |
| **Actors** | Admin users |
| **Trigger** | User clicks either the "print" or the "save" button |
| **Pre-conditions** | 1. Authenticated session.<br>2. Target information is selected. |
| **Post-conditions** | 1. Target information is outputted to the appropriate media. |
| **Main Case** | 1. Admin user clicks on the "print" or the "save" button from the page.<br>2. System outputs the information as requested. |
| **Exception Path** | 1. User did not select any information.<br>2. System selects all the information on display and proceeds with the request.<br><br>1. Database error.<br>2. System logs the error along with the information on display.<br>3. System notifies user of unsuccessful update and requests user to retry. |

Table 5.4.4 Print/Save Audited Logs.

## 5.5 Use Cases for Error Management

### 5.5.1 List Error Messages

| ***List Error Messages*** | |
|---|---|
| **Description** | Displays a list of error messages |
| **Actors** | System administrators |
| **Trigger** | User selects the "Errors" menu or the "Errors" option from the main page. |
| **Pre-conditions** | 1. Authenticated session. |
| **Post-conditions** | 1. A list of errors are displayed |
| **Main Case** | 1. Admin user clicks on the "Errors" option from either the menu or the main page.<br>2. System outputs the information, filtered according to user's permission level.<br>3. Admin user may choose to sort the options by a given field, or to search for a particular error message. If this is the case, system responds by narrowing the list of error messages displayed. |
| **Exception Path** | 1. Database error.<br>2. System logs the error along with the information on display.<br>3. System notifies user of unsuccessful update and requests user to retry. |

Table 5.5.1 List Error Messages.

### 5.5.2 Print/Save Error Messages

| ***Print/Save Error Messages*** | |
|---|---|
| **Description** | Outputs the error messages as requested. |
| **Actors** | System administrators |
| **Trigger** | User clicks either the "print" or the "save" button. |
| **Pre-conditions** | 1. Authenticated session.<br>2. Target error messages are selected. |
| **Post-conditions** | 1. Target information is outputted to the appropriate media. |
| **Main Case** | 1. Admin user clicks on the "print" or the "save" button from the page.<br>2. System outputs the detailed information of the item(s) selected to the requested media. |
| **Exception Path** | 1.1. User did not select any information.<br>1.2. System selects all the information on display and proceeds with the request.<br><br>2.1. Database error.<br>2.2. System logs the error along with the information on display.<br>2.3. System notifies user of unsuccessful update and requests user to retry. |

Table 5.5.2 Print/Save Error Messages.

### 5.5.3 View More Details/Annotate Error Messages

| ***View More Details/Annotate Error Messages*** | |
|---|---|
| **Description** | Displays the detailed description of the error messages, and allows user to annotate as appropriate. |
| **Actors** | System administrators |
| **Trigger** | User clicks the "More details/Edit" button. |
| **Pre-conditions** | 1. Authenticated session.<br>2. Target information is selected. |
| **Post-conditions** | 1. Target information is displayed, and an additional field for comments is shown. |
| **Main Case** | 1. Admin user selects the error message(s) of interest and clicks on the "More details" button.<br>2. System outputs the information as requested.<br>3. Admin user may print the detailed information (as described in use case 5.4.2) or add a comment (for instance, the steps taken to resolve the issue).<br>4. Admin user may choose to save the comments, or to exit without saving changes. If this is the case, then system returns admin user to the list of error messages. |
| **Exception Path** | 1.1. User did not select any information.<br>1.2. System selects all the information on display and proceeds with the request.<br><br>2.1. Database error.<br>2.2. System logs the error along with the information on display.<br>2.3. System notifies user of unsuccessful update and requests user to retry. |

Table 5.5.3 View More Details/Annotate Error Messages.

## 5.6 Use Cases for Request Management

### 5.6.1 Use Case to View Pending Requests

| ***View Pending Requests*** | |
|---|---|
| **Description** | Displays pending requests submitted by users with lesser permissions than admin user. |
| **Actors** | Users with permission levels 1, 2 or 3 |
| **Trigger** | User selects "Requests" from the menu. |
| **Pre-conditions** | User is logged in |
| **Post-conditions** | Viewing pending requests page |
| **Main Case** | 1. System retrieves the pending requests allowed to be viewed by user<br>2. System displays the pending requests retrieved |
| **Exception Path** | 1. Database error<br>2. System displays an error message |

Table 5.6.1 Use Case to View Pending Requests. **Note:** User may view the requests made by users of the lower level(s) as well as by themselves.

### 5.6.2 Use Case to Approve a Request

| ***Approve Request*** | |
|---|---|
| **Description** | User approves a pending request. |
| **Actors** | Users with permission level 1, 2 or 3 |
| **Trigger** | User clicks the "Approve/Formalize request" button |
| **Pre-conditions** | 1. User is logged in.<br>2. User is viewing pending requests. |
| **Post-conditions** | 1. The request is formalized.<br>2. The request approved, or the request is re-submitted at the permission level of user. |
| **Main Case** | 1. System displays the request page pre-populated with the requester's name, and currently entered fields.<br>2. User completes the page as required (e.g. by adding the information corresponding to the description of the request).<br>3. User selects "approve" button.<br>4. System asks user to confirm the approval.<br>5. System saves the request and displays a "Request approval/formalization successful" message. |
| **Exception Path** | 1.1. Incomplete information in page (for instance, no building entered in a transfer request)<br>1.2. System displays an error message describing the missing fields.<br>1.3. User is given the opportunity to re-edit the form.<br><br>2.1. A database error occurs after the request approval/formalization is submitted by user.<br>2.2. System logs the error along with the details of the request approval/formalization.<br>2.3. System requests user to contact a system administrator. |

Table 5.6.2 Use Case to Approve a Request. **Note:** When a user wishes to approve a request, he/she should formalize the request. Formalizing a request entails translating the textual description into the proper identification numbers by optionally using the search function. It may be possible for the request to be elevated in permission level as part of the approval procedure. In this case, the "approval" simply submits the request to the proper level.

### 5.6.3 Use Case to Reject a Request

| ***Reject Request*** | |
|---|---|
| **Description** | User requesting to "reject request" |
| **Actors** | Users with permission level 1, 2 or 3 |
| **Trigger** | User clicks the "reject" button |
| **Pre-conditions** | 1. User is logged in<br>2. User is able to view pending requests |
| **Post-conditions** | The request is rejected. |
| **Main Case** | 1. System displays a page containing a brief summary of the request and a "comment box".<br>2. User may fill in the comment box to detail the reason(s) for rejecting the request.<br>3. User selects reject button.<br>4. System asks user to confirm the rejection.<br>5. User confirms the rejection.<br>6. System updates the status of the request.<br>7. System displays a message indicating the successful rejection of the request.<br>8. System notifies user of the rejection decision. |
| **Exception Path** | 1. A database error occurs while the request is processed.<br>2. System logs the error along with the details of the rejection.<br>3. System locks the request.<br>4. System requests user to contact a system administrator. |

Table 5.6.3 Use Case to Reject a Request.

## 6. Data Dictionary [8]

Table 6.1 ACL

| Data Member Name | Description | Type | Additional Type Information | Default Value | Mandatory? | Unique ? |
|---|---|---|---|---|---|---|
| user_role_id | User identification | Integer | 3 followed by a nine-digit number | Null | Yes | Yes |
| permission | Permissions (bit-sum of allowed actions) | Integer | Up to 65535 | Null | Yes | No |

Table 6.1 ACL. Contains permissions per user. Permissions are the compositions of bits. Compositions are set up where a role is granted and can be changed by administrators.

Table 6.2 Buildings

| Data Member Name | Description | Type | Additional Type Information | Default Value | Mandatory? | Unique ? |
|---|---|---|---|---|---|---|
| bldg_id | Building identification | Integer | 0 followed by a nine-digit number | Null | Yes | Yes |
| bldg_code | Abbreviation or short name | Varchar | 10 | Null | Yes | Yes |
| bldg_name | Name of the building | Varchar | 50 | Null | Yes | Yes |

Table 6.2 Buildings. Stores the information of buildings.

Table 6.3 Item Categories

| Data Member Name | Description | Type | Additional Type Information | Default Value | Mandatory? | Unique ? |
|---|---|---|---|---|---|---|
| cat_id | Identification of category | Integer | 5 followed by a nine-digit number | Null | Yes | Yes |
| parent_cat_id | Parent identification of category | Integer | 5 followed by a nine-digit number | Null | No | No |
| description | Description of the category | Varchar | 50 | Null | Yes | No |

Table 6.3 Item Categories. Contains information of categories. Categories are divided in 2 levels. Level 1 categories contain be broad categories for which parent_cat_id should be 0. Level 2 categories are the subcategories in which parent_cat_id should be assigned to a cat_id corresponding to a Level 1 category.

Table 6.4 Affiliations

| Data Member Name | Description | Type | Additional Type Information | Default Value | Mandatory? | Unique ? |
|---|---|---|---|---|---|---|
| affln_id | Identification of department or of faculty | Integer | 1 followed by a nine-digit number | Null | Yes | Yes |
| affln_name | Name of department / faculty | Varchar | 50 | Null | Yes | Yes |
| alffln_code | Abbreviation of department / faculty | Varchar | 10 | Null | Yes | Yes |

Table 6.4 Affiliations. Contains information describing the organization of the university. There are 3 levels: Level 1 is for the university, level 2 is for faculties and level 3 for departments. Level 1 is represented by nine digits of "zero". The first two (non-zero) digits are reserved to denote the faculty, and faculties themselves contain two non-zero digits followed by a string of seven digits of "zero". The ID for departments contain the two first digits showing the parent faculty followed by seven (non-zero) digits representing the department.

Table 6.5 Items

| Data Member Name | Description | Type | Additional Type Information | Default Value | Mandatory? | Unique ? |
|---|---|---|---|---|---|---|
| item_id | Identification of item | Integer | 4 followed by a nine-digit number | Null | Yes | Yes |
| item_description | Description of item | Varchar | 50 | Null | Yes | No |
| code | Tracking code | Varchar | 20 | Null | Yes | Yes |
| group_id | Group ID | Integer | 5 followed by a nine-digit number | Null | No | No |
| serial_number | Product serial number | Varchar | 20 | Null | Yes | Yes |
| cat_id | Category identification | Integer | 4 followed by a nine-digit number | Null | Yes | No |
| owner_id | ID of owner (member, department / faculty, university) | Integer | Ten-digit number | 0000000000 | Yes | No |
| loc_id | Location Id | Integer | 0 followed by a nine-digit number | Null | Yes | No |
| date_modified | Timestamp of modification | Timestamps | N/A | N/A | Yes | No |
| status | Status of the item | Varchar | Ten-character string | Null | No | No |

Table 6.5 Items. Contains information of specific items, one per record.

Table 6.6 Item Property List

| Data Member Name | Description | Type | Additional Type Information | Default Value | Manda tory? | Unique ? |
|---|---|---|---|---|---|---|
| prop_id | Identification of the property | Integer | 5 followed by a nine-digit number | Null | Yes | Yes |
| cat_id | Identification of the item type for which the property is relevant | Integer | 5 followed by a nine-digit number | Null | Yes | No |
| prop_name | Name of the property | Varchar | Ten-character string | Null | Yes | No |
| default_valu e | Default value for the property | Varchar | Twenty-character string | Null | No | No |

Table 6.6 Item Property List. Lists the properties for each item type (for instance, computer mice may have a "detection method" to indicate whether the mouse is a laser, optical or wheel mouse, and a biological mouse may have a "genotype" property).

Table 6.7 Item Properties

| Data Member Name | Description | Type | Additional Type Information | Default Value | Manda tory? | Unique ? |
|---|---|---|---|---|---|---|
| item_prop_i d | Identification of the item/ property pair | Integer | 6 followed by a nine-digit number | Null | Yes | Yes |
| item_id | Identification of the item | Integer | 4 followed by a nine-digit number | Null | Yes | No |
| prop_id | Identification of the property | Integer | 5 followed by a nine-digit number | Null | Yes | No |
| prop_value | Value of the property | Varchar | Twenty-character string | Null | Yes | No |

Table 6.7 Item Properties. Lists all the relevant properties of each item.

Table 6.8 Locations

| Data Member Name | Description | Type | Additional Type Information | Default Value | Manda tory? | Unique ? |
|---|---|---|---|---|---|---|
| loc_id | Identification of location | Integer | 0 followed by a nine-digit number | Null | Yes | Yes |
| parent_loc_ id | Parent identification of location | Integer | 0 followed by a nine-digit number | Null | Yes | No |
| loc_code | Short name or abbreviation | Varchar | 10 | Null | Yes | yes |
| loc_name | Name of location | Varchar | 50 | Null | Yes | No |
| bldg_id | Identification of building or container location (floor, room, etc.) | Integer | 0 followed by a nine-digit number | Null | Yes | No |
| affln_id | Owner of the location | Integer | Ten-digit number | Null | Yes | No |
| Status | status of the location (available, booked, in-use, etc.) | Varchar | 10 | "Available" | Yes | No |
| loc_type_id | Reference to Location type | Integer | Ten-digit number | Null | Yes | No |
| Comment | Comment on the location | Varchar | 255 | Null | No | No |

Table 6.8 Items. Contains information describing physical locations. A location may contain other locations.

Table 6.9 Location Types

| Data Member Name | Description | Type | Additional Type Information | Default Value | Manda tory? | Unique ? |
|---|---|---|---|---|---|---|
| loc_type_id | Identification of location | Integer | Ten-digit number | Null | Yes | Yes |
| loc_type_n ame | English name for type of location | Varchar | 15 | Null | Yes | Yes |
| Description | Description of location | Varchar | 255 | Null | Yes | No |

Table 6.9 Location Types. Contains information describing the different types of location.

Table 6.10 Permissions

| Data Member Name | Description | Type | Additional Type Information | Default Value | Manda tory? | Unique ? |
|---|---|---|---|---|---|---|
| permission _id | Identification of permission | Integer | Up to 16 | Null | Yes | Yes |
| Description | description of permission | Varchar | 255 | Null | Yes | No |

Table 6.10 permissions. Contains definitions of each permittion. Each permission takes one bit of a big integer.

Table 6.11 Requests

| Data Member Name | Description | Type | Additional Type Information | Default Value | Mandatory? | Unique ? |
|---|---|---|---|---|---|---|
| req_id | Identification of request | Integer | Ten-digit number | Null | Yes | Yes |
| requester | Id of the party who will receive the requested item (typically the same of submitted_by) | Integer | Ten-digit number | Null | No | No |
| req_type | Request type | Integer | Up to 8 | Null | Yes | No |
| submitted_by | Id of party who issues the request | Integer | Ten-digit number | Null | Yes | No |
| item_id | Item/member/ location id requested for | Integer | Ten-digit number | Null | No | Yes |
| description | Description of the request, also used for reporting a problem | Varchar | 255 | Null | Yes | No |
| date_submitted | Date/time of submission | Datetime | N/A | Null | Yes | No |
| approved_by | ID of party who approved or disapproved the requester | Integer | Ten-digit number | Null | No | No |
| date_approved | Datetime of approval | Datetime | N/A | Null | No | No |
| Status | Indication the status of the request | Varchar | 10 | "InProcess" | Yes | No |
| date_modified | Timestamp of modification | Timestamp | N/A | Null | No | No |

Table 6.11 Requests. Contains the information of requests. A request may be issued by user his/herself, or delegated by upper level users. Normal users may only view his/her requests (requester's ID is his/her own). The request_type is used for distinguishing the requests such as check out an item or report a problem.

Table 6.12 Request Types

| Data Member Name | Description | Type | Additional Type Information | Default Value | Mandatory? | Unique ? |
|---|---|---|---|---|---|---|
| req_type_id | Identification of request type | Integer | 2 followed by a nine-digit number | Null | Yes | Yes |
| req_type_code | Abbreviation describing the type of request | Varchar | 10 | Null | Yes | Yes |
| description | Name of type of request, and its description | Varchar | 10 | Null | No | No |
| permission | The permissions required to treat this request | Integer | Up to 65535 | 0 | Yes | No |

Table 6.12 Request Types. Describes the different types of requests, as well as the permissions required to handle the requests of a given type.

Table 6.13 Professional Titles

| Data Member Name | Description | Type | Additional Type Information | Default Value | Manda tory? | Unique ? |
|---|---|---|---|---|---|---|
| title_id | Identification of the professional title of a user | Integer | 2 followed by a nine-digit number | Null | Yes | Yes |
| title_name | Role name | Varchar | 50 | Null | Yes | No |
| permission | The default permissions of the role | Integer | Up to 65535 | 0 | Yes | No |

Table 6.13 Professional Titles. Contains the predefined definition of each role and the default permissions. Permissions are the compositions of bits.

Table 6.14 Users

| Data Member Name | Description | Type | Additional Type Information | Default Value | Manda tory? | Unique ? |
|---|---|---|---|---|---|---|
| user_id | User's identification | Integer | 2 followed by a nine-digit number | Null | Yes | Yes |
| user_code | Student username (e.g. for login) | Varchar | Ten-character string | Null | Yes | Yes |
| last_name | Last name | Varchar | 20 | Null | Yes | No |
| first_name | First name | Varchar | 20 | Null | Yes | No |
| password | Password | Varchar | 50 | Null | Yes | No |
| date_modif ied | Timestamps for modification | Time stamps | N/A | Null | No | No |
| login_atte mpts | Number of consecutive failed attempts | Integer | Byte | 0 | Yes | No |

Table 6.14 Users. Contains the user's basic information. Password field holds encryped password.

Table 6.15 User Info

| Data Member Name | Description | Type | Additional Type Information | Default Value | Mandatory? | Unique ? |
|---|---|---|---|---|---|---|
| user_id | User's indentification | Integer | 2 followed by a nine-digit number | Null | Yes | Yes |
| email | Email address | Varchar | 255 | Null | Yes | No |
| dob | Date of birth | Date | N/A | Null | Yes | No |
| home_phone | Telephone number | Integer | Ten-digit number | Null | No | No |
| cell_phone | Cell phone number | Integer | Ten-digit number | Null | No | No |
| street_address | Mailing address | Varchar | 255 | Null | Yes | No |

Table 6.15 User Info. Contains additional user information.

Table 6.16 User Roles

| Data Member Name | Description | Type | Additional Type Information | Default Value | Mandatory? | Unique? |
|---|---|---|---|---|---|---|
| user_role_id | Identification of user/role combination | Integer | 3 followed by a nine-digit number | Null | Yes | Yes |
| user_id | Identification of user. | Integer | 2 followed by a nine-digit number | Null | Yes | No |
| title_id | Identification of role | Integer | 2 followed by a nine-digit number | Null | Yes | No |
| affln_id | Identification of department | Integer | 1 followed by a nine-digit number | Null | Yes | No |
| status | Status of the role (accepted, dropped, etc.) | Varchar | 10 | Null | Yes | No |

Table 6.16 User Roles. Contains the user roles for users. A single user may be granted more than one role.

Table 6.17 Inventories

| Data Member Name | Description | Type | Additional Type Information | Default Value | Mandatory? | Unique ? |
|---|---|---|---|---|---|---|
| item_id | Identification of item | Integer | 4 followed by a nine-digit number | Null | Yes | Yes |
| qty | The available quantity of item | Integer | Up to 65535 | 1 | Yes | No |
| status | The status of the item | Varchar | 10 | "Available" | Yes | No |
| modified_by | Who modified the record | Integer | Ten-digit number | Null | Yes | No |
| date_modified | Timestamps | Time stamps | N/A | Null | Yes | No |

Table 6.17 Inventories. Contains the inventory information. This purpose of this inventory is to allow future implementations of a lending system, rather than simply list all items (as in Table 6.5). If an item is checked out, the quantity will be decreased by 1. If the quantity becomes 0, the status will be updated to "Checked out".

Table 6.18 Table List

| Data Member Name | Description | Type | Additional Type Information | Default Value | Mandatory? | Unique ? |
|---|---|---|---|---|---|---|
| table_id | Identification number of table | Integer | Up to 255 | Null | Yes | Yes |
| table_code | Database name of table | Varchar | Up to 10 characters | Null | Yes | Yes |
| table_name | User-friendly description of table | Varchar | Up to 15 characters | Null | Yes | Yes |
| permissions | Minimum permission signature for table | Integer | Up to 65535 | Null | Yes | No |

Table 6.18 Table List. Lists all tables (except for those described in Table 6.18 and Table 6.19) in the database system of the UUIS. This table serves to enumerate the possibilities for searching, but is otherwise irrelevant to the UUIS.

Table 6.19 Field List

| Data Member Name | Description | Type | Additional Type Information | Default Value | Mandatory? | Unique ? |
|---|---|---|---|---|---|---|
| field_id | Identification number of field | Integer | Up to 255 | Null | Yes | Yes |
| table_id | Identification of parent table | Integer | Up to 255 | Null | Yes | No |
| field_code | Database name of field | Varchar | Up to 15 | Null | Yes | No |
| field_name | User-friendly description of field | Varchar | Up to 20 characters | Null | Yes | No |
| permissions | Minimum permission signature for table | Integer | Up to 65525 | Null | Yes | No |

Table 6.19 Field List. Lists all the fields from all tables (except for those described in Table 6.18 and Table 6.19) in the database system of the UUIS. This table serves to enumerate the possibilities for searching, but is otherwise irrelevant to the UUIS.

Table 6.20 Logs

| Data Member Name | Description | Type | Additional Type Information | Default Value | Manda tory? | Unique ? |
|---|---|---|---|---|---|---|
| log_id | Identification of log | Integer | Up to 4 bytes | Null | Yes | Yes |
| log_time | Date time when the log recorded | datetime | N/A | Null | Yes | No |
| user_id | Who responsible for the event | Integer | 3 followed by a nine-digit number | Null | Yes | No |
| item_id | “Item” of interest (location, user, etc.) | Integer | Ten-digit number | Null | Yes | No |
| event_type | Event type | Varchar | 10 | Null | Yes | No |
| content | Content of the event | Varchar | 255 | Null | Yes | No |

Table 6.20 Logs. Contains the information of system activity, automatically filled and only viewable when auditing.

## 7. Mock User Interface Screenshots

In this section, we will present selected screenshots in order to provide a preliminary presentation of the design for the user interface.

### 7.1 Login

In this page, the user will be requested to input his/her username and his/her password.

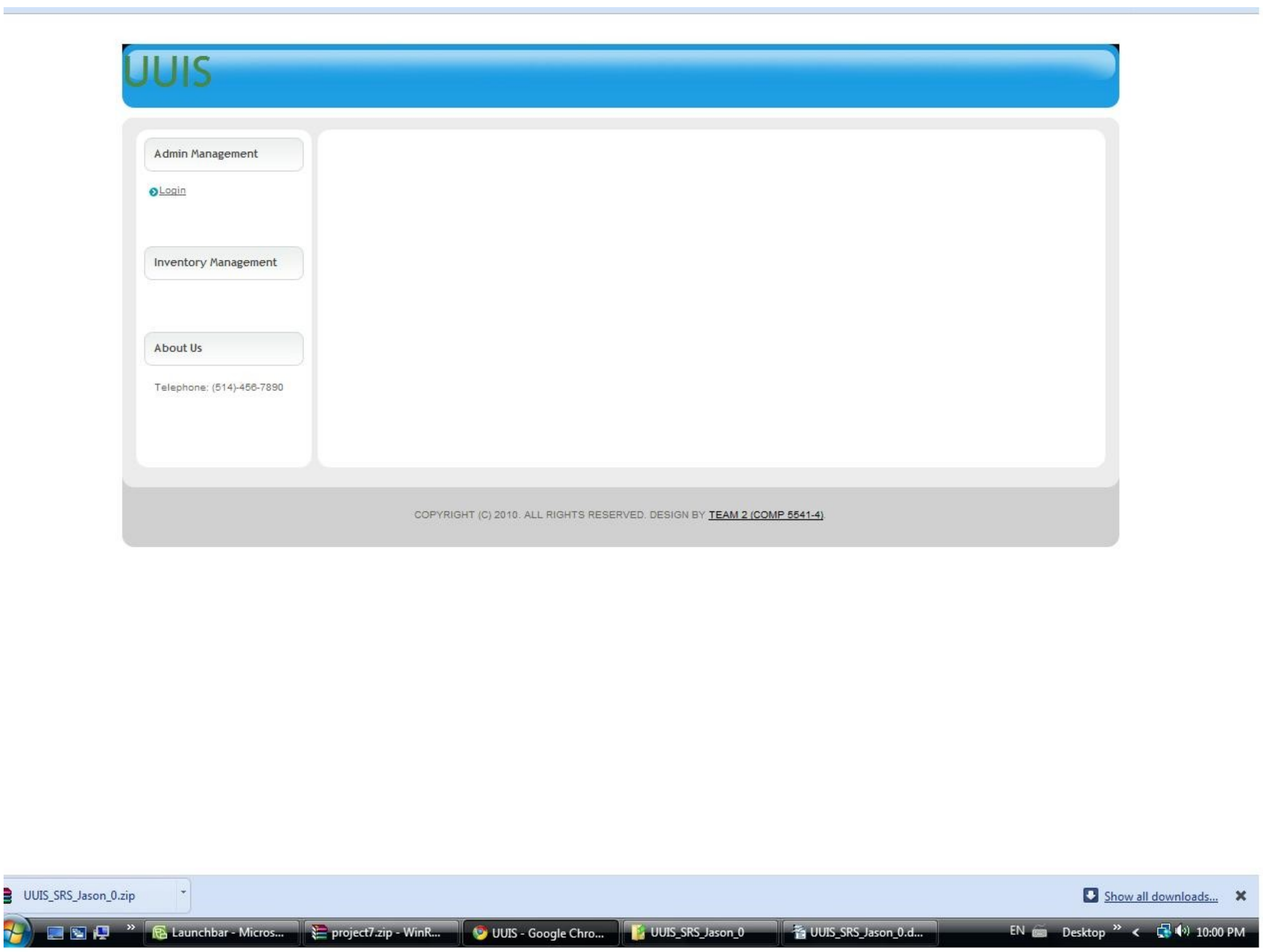

Fig. 7.1 Login Page

## 7.2 Welcome Screen

In this page, the user will be presented on the left-hand side with the different options available to the user for interaction with the UUIS, as well as a set of criteria for sorting/searching the items in the database.

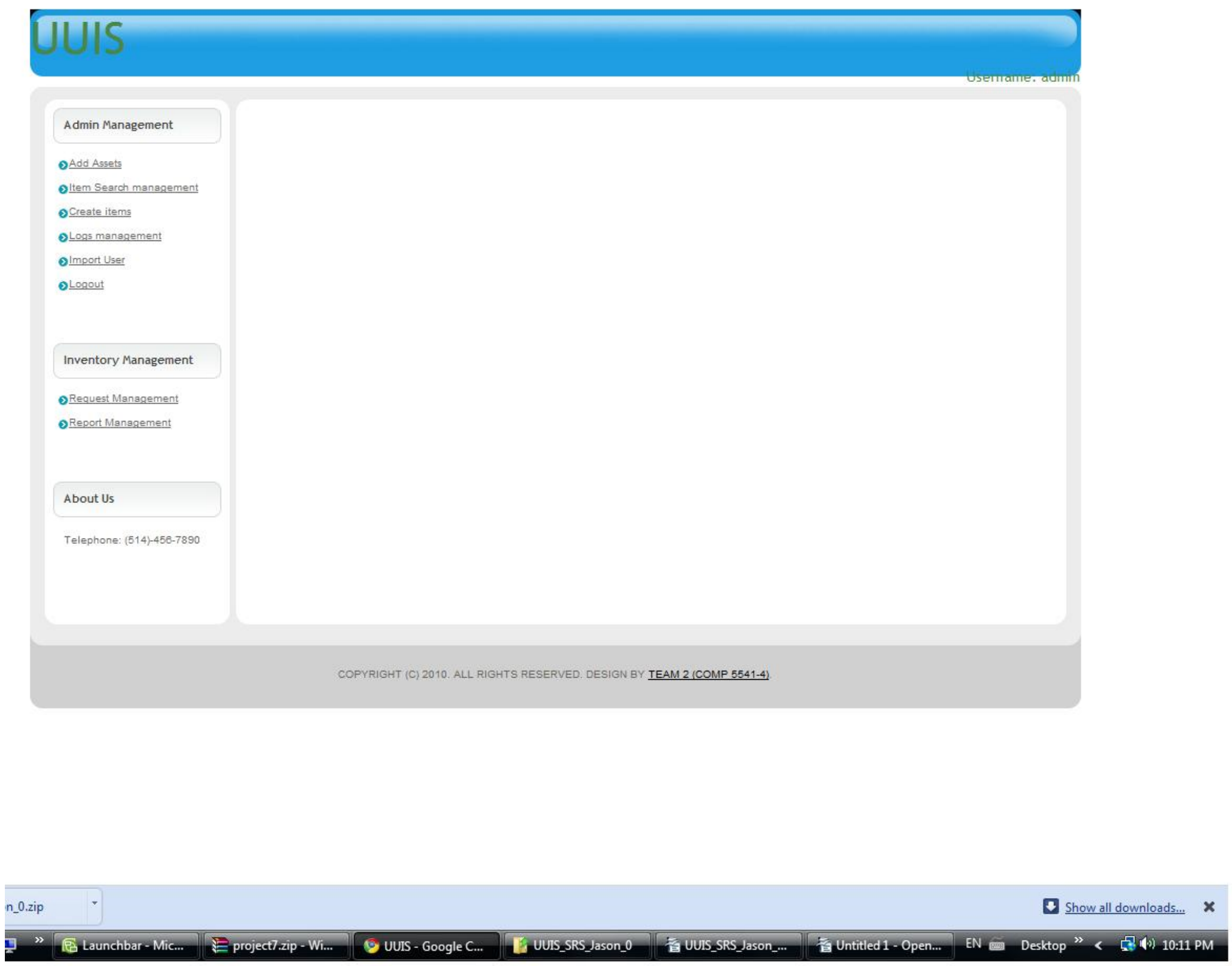

Figure 7.2 Welcome Screen

### 7.3 Search Page

In this page, the user may search criteria desired from text boxes. In addition, we will also provide a textbox advanced search with “and”, “not” or “or” options.

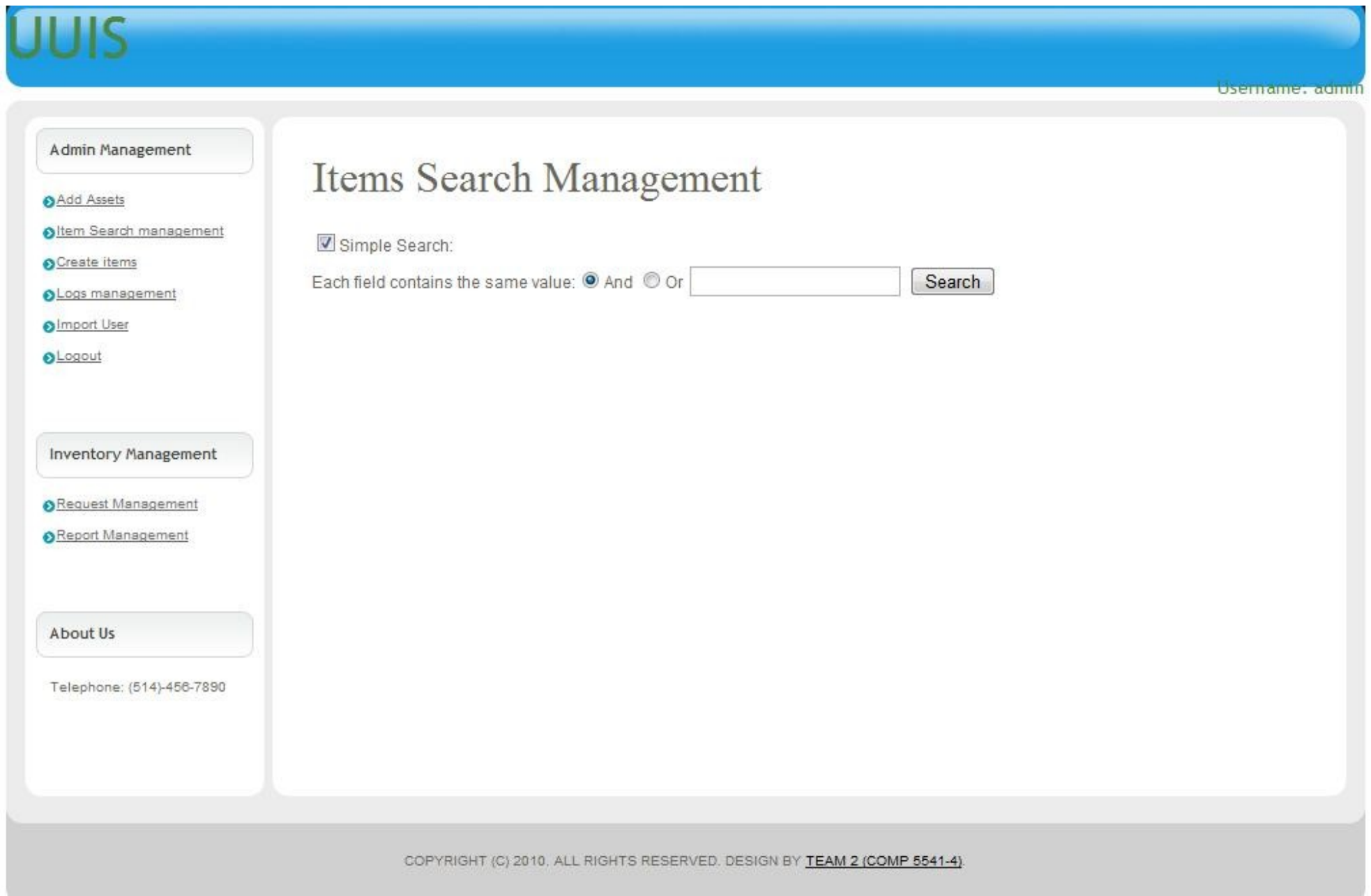

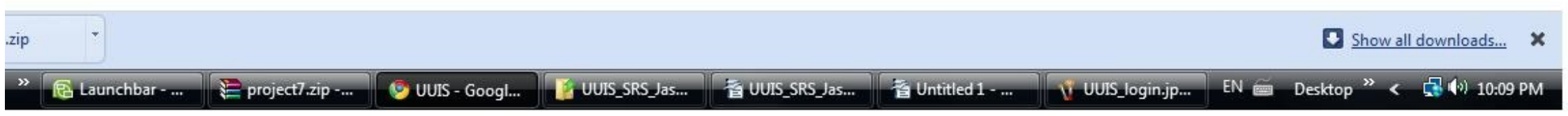

Figure 7.3.1 Basic Search Page

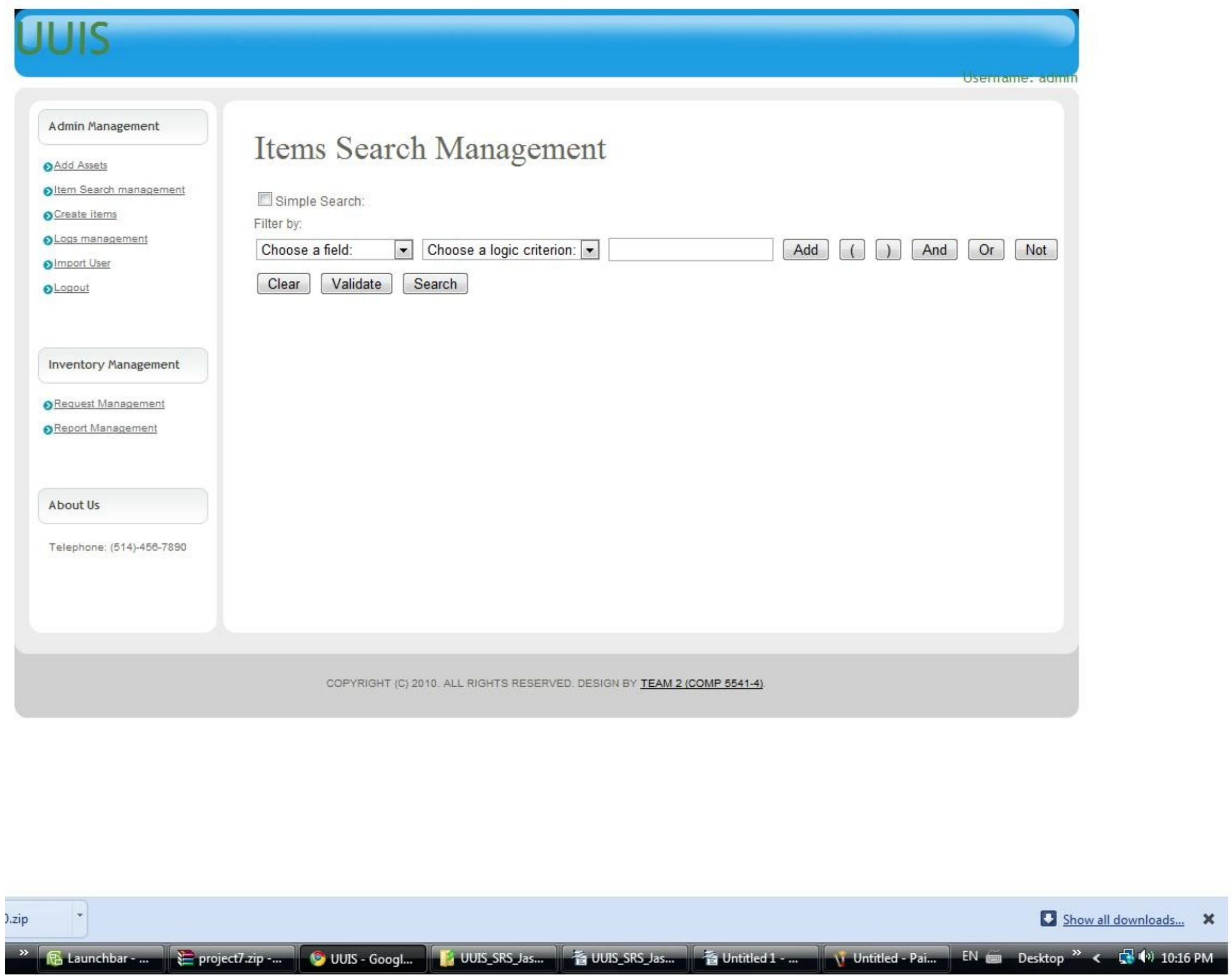

Figure 7.3.2 Advanced Search Page

## 8. Cost Estimation[23]

**Time Estimation**

(in hours)

| Methodology Stage or Activity | Best Case | Most likely Case | Worst Case | Estimated Case | Standard Deviation | Task Estimate 95% Confidence |
|---|---|---|---|---|---|---|
| **Requirement** | 34.00 | 60.00 | 85.00 | 59.90 | 8.50 | 76.90 |
| Introduction | 4.00 | 8.00 | 12.00 | 8.00 | 1.40 | 10.80 |
| Project Background | 6.00 | 10.00 | 14.00 | 10.00 | 1.40 | 12.80 |
| Project Rationale | 4.00 | 10.00 | 14.00 | 9.70 | 1.70 | 13.10 |
| Primary Research | 20.00 | 32.00 | 45.00 | 32.20 | 4.20 | 40.60 |
| **Analysis** | 72.00 | 126.00 | 182.00 | 126.40 | 18.40 | 163.20 |
| Evaluation of Primary Research | 30.00 | 50.00 | 70.00 | 50.00 | 6.70 | 63.40 |
| Secondary Research | 15.00 | 24.00 | 35.00 | 24.40 | 3.40 | 31.20 |
| Feasibility Study | 12.00 | 22.00 | 32.00 | 22.00 | 3.40 | 28.80 |
| Risk Management | 15.00 | 30.00 | 45.00 | 30.00 | 5.00 | 40.00 |
| **Design** | 105.00 | 194.00 | 350.00 | 205.20 | 40.90 | 287.00 |
| Logical Design | 20.00 | 40.00 | 70.00 | 41.70 | 8.40 | 58.50 |
| UML | 9.00 | 22.00 | 42.00 | 23.20 | 5.50 | 34.20 |
| Use Cases | 4.00 | 10.00 | 20.00 | 10.70 | 2.70 | 16.10 |
| Class Diagram | 5.00 | 12.00 | 22.00 | 12.50 | 2.90 | 18.30 |
| Solution Sketch | 22.00 | 40.00 | 70.00 | 42.00 | 8.00 | 58.00 |
| Data Flow Diagram | 2.00 | 6.00 | 10.00 | 6.00 | 1.40 | 8.80 |
| Database Design | 50.00 | 80.00 | 150.00 | 86.70 | 16.70 | 120.10 |
| Entity Relation Ship Diagram | 2.00 | 6.00 | 8.00 | 5.70 | 1.00 | 7.70 |
| **Coding** | 400.00 | 500.00 | 949.00 | 558.20 | 91.50 | 741.20 |
| Learning Technology | 60.00 | 70.00 | 150.00 | 81.70 | 15.00 | 111.70 |
| Physical Design | 20.00 | 40.00 | 75.00 | 42.50 | 9.20 | 60.90 |
| Development | 320.00 | 390.00 | 724.00 | 434.00 | 67.40 | 568.80 |
| Database Development | 20.00 | 50.00 | 70.00 | 48.40 | 8.40 | 65.20 |
| Coding | 300.00 | 340.00 | 654.00 | 385.70 | 59.00 | 503.70 |
| **Testing** | 12.00 | 22.00 | 44.00 | 24.00 | 5.40 | 34.80 |
| Test Cases | 12.00 | 22.00 | 44.00 | 24.00 | 5.40 | 34.80 |
| Individual Module Testing | 4.00 | 8.00 | 1 6.00 | 8.70 | 2.00 | 12.70 |
| Integration Testing | 5.00 | 8.00 | 1<br>5.00 | 8.70 | 1.70 | 12.10 |
| Performance Testing | 2.00 | 4.00 | 8.<br>00 | 4.40 | 1.00 | 6.40 |
| Acceptance Testing | 1.00 | 2.00 | 5.<br>00 | 2.40 | 0.70 | 3.80 |
| **Deployment** | 67.00 | 92.00 | 160.00 | 99.20 | 15.50 | 130.20 |
| Deployment Plan | 4.00 | 6.00 | 10.00 | 6.40 | 1.00 | 8.40 |
| Critical Information | 2.00 | 4.00 | 7.00 | 4.20 | 0.90 | 6.00 |
| Documentation | 60.00 | 80.00 | 140.00 | 86.70 | 13.40 | 113.50 |
| Conclusion | 1.00 | 2.00 | 3.00 | 2.00 | 0.40 | 2.80 |

| **Group Meetings:** | **Estimated Time** |
|---|---|
| Face to Face meetings | 80.00 |
| Online Group meetings | 180.00 |
| Total cost | 7800.00 |

| **Total:** | |
|---|---|
| - | |
| Estimated Case (Project Work) | 1073.00 |
| Standard Deviation | 14 |
| Project Estimate 95% confidence | 1101 |
| Hourly Rate: $ 30.00 | |
| Group Meetings: | 7800.00 |
| Total Cost: | 39990 $ |

Table 8.1 Cost Estimation. Details the estimated cost of this project.